\documentclass{camera}
\usepackage{epsfig}

\begin{document}

%
\title{Neutrino physics from Cosmology}

%
\author{Steen Hannestad}

%
\organization{Department of Physics and Astronomy, Aarhus University, Ny Munkegade, DK-8000 Aarhus C, Denmark}

\maketitle

\begin{abstract}
In recent years precision cosmology has become an increasingly powerful probe of particle
physics. Perhaps the prime example of this is the very stringent cosmological upper bound on the
neutrino mass. However, other aspects of neutrino physics, such as their decoupling history
and possible non-standard interactions, can also be probed using observations of cosmic structure. Here,
I review the current status of cosmological bounds on neutrino properties and discuss the
potential of future observations, for example by the recently approved EUCLID mission, to
precisely measure neutrino properties.
\end{abstract}

%
\section{Introduction}

The past decade has clearly established cosmology as a very powerful laboratory for neutrino physics, capable of probing neutrino
properties which are inaccessible to terrestrial experiments. Historically, light element formation in early Universe, also
known as Big Bang nucleosynthesis, was used to probe neutrino interactions beyond the standard model as well as the relativistic
energy density in neutrinos around the time where neutrinos decouple from thermal equilibrium with the electromagnetically 
interacting plasma.
More recently attention has shifted to late time cosmology, and in particular to the effect of neutrinos on how large scale structure
forms. Even though neutrinos are a very subdominant component of the cosmological energy density at present they have a profound impact on structure formation because the anisotropic stress caused by neutrino free-streaming leads to a strong suppression of structure formation growth. 

Here, I will review the current status of cosmological bounds on neutrino properties (see e.g.\ \cite{Hannestad:2010kz,wong,lesgourgues} for more detailed and recent reviews of
neutrino cosmology), both within the standard model and in various extensions of it.

\section{Cosmological bounds on standard model neutrinos}

In the standard model the influence of neutrinos on structure formation can be described in terms of just one parameter, the sum of neutrino masses, $\sum m_\nu$. The reason is that structure formation is mainly influenced by the neutrino contribution to the cosmic energy density, $\Omega$. The number density of neutrinos is precisely calculable in the standard model and is almost identical for all neutrino mass states. To a reasonable approximation neutrinos decouple in the early Universe at a temperature of approximately 2-3 MeV, shortly before electrons and positrons annihilate. The entropy transfer to photons induced by this process leads to the prediction that the neutrino temperature is related to the photon temperature by $T_\nu \simeq (4/11)^{1/3} T_\gamma$. From this temperature the number density of neutrinos can be calculated and is $n_\nu = 3/4 (T_\nu/T_\gamma)^3 n_\gamma$ for each species of neutrino.
Since $\Omega_\nu h^2 = \sum n_{\nu,i} m_{\nu,i} = n_\nu \sum m_{\nu,i} = n_\nu \sum m_\nu$ the contribution of neutrinos to $\Omega$ can be cast as $\Omega_\nu h^2 = \sum m_\nu/94.6$ eV. Any cosmological constraint on $\Omega_\nu h^2$is therefore directly translatable into a bound on the sum of neutrino masses.
The most robust cosmological bounds come from measurements of the cosmic microwave background by the Planck satellite and are $\sum m_\nu < 0.933$ eV for Planck data only \cite{Ade:2013zuv} (but including polarization data from the WMAP satellite), strengthening to $m_\nu < 0.24$ eV when auxiliary data on baryon acoustic oscillations is included. However, it should be stressed here that this limit is derived for a 1-parameter extension to the vanilla $\Lambda$CDM model. For more complex cosmological models the bound on neutrino masses can be significantly weakened.

\section{Beyond the standard model}

Beyond the standard model a plethora of new possibilities opens up. As described above the temperature of a single species of standard model neutrinos is approximately given by $T_\nu \simeq (4/11)^{1/3} T_\gamma$ and for historical reasons one unit of neutrino energy density in the early Universe is given by the energy density in a single spin state of a fermion with exactly this temperature. The first prediction would therefore be that $\rho_\nu = N_\nu \rho_{\nu,0}$ with $N_\nu=3$. However, the definition of $\rho_{\nu,0}$ ignores the effect of $e^+e^-$annihilation to neutrinos and finite temperature corrections to the photon propagator, and the correct standard model prediction is $N_\nu=3.046$.

While the standard model predicts exactly this value various types of physics beyond the standard model can change this number. This can happen for a number of different reasons, some of them related to neutrino physics and others to the cosmological model.

If the modification to $N_\nu$comes from neutrino physics it can for example be caused by the presence of a non-zero lepton asymmetry in the neutrino sector or by the presence of additional light neutrino species - a possibility I will discuss further in the folowing subsection. However, observations of cosmic structure ofrmation cannot in themselves distinguish energy density in neutrinos from energy density in other light species and therefore $N_\nu$ should more appropriately be thought of as quantifying the amount of non-electromagnetically interacting relativistic energy density, also known as {\it dark radiation}.

The current bound from the Planck satellite data combined with WMAP polarization data is $N_\nu = 3.51^{+0.80}_{-0.74}$ \cite{Ade:2013zuv}, assuming a vanilla $\Lambda$CDM+$N_\nu$ model. However, while this is perfectly consistent with the standard model prediction the inferred value of the Hubble parameter comes out significantly lower than the direct measurement of the local $H_0$ value. Adding the locally measured $H_0$ as a prior shifts the preferred $N_\nu$ to a higher value, $N_\nu = $, possibly making it inconsistent with the standard model prediction. In addition to this the bound on $N_\nu$ can also change as soon as additional model parameters are used. For example models with non-zero neutrino masses, either in the active, or in the sterile sector, typically prefer values of $N_\nu$ significantly higher than 3.046 when fitted to current data (see e.g.\ \cite{Hamann:2013iba,Archidiacono:2013fha}). In summary it is at present unclear whether there is tension between the predicted and the measured value of $N_\nu$. 
It should also be stressed here that in fact $N_\nu$ different from 3.046 can be caused by changes to the cosmological model rather than by dark radiation. 

In addition to modifying $N_\nu$  there are plenty of other possiblities for beyond standard model physics in the neutrino sector: New neutrino species, non-standard neutrino interactions, etc. Here I will simply review two often discussed extensions of standard model neutrino physics: sterile neutrinos and non-standard interactions.

\subsection{Sterile neutrinos}

Sterile neutrino can in the present context be thought of as additional fermions which are singlets under the standard model gauge group, but have mixing with the active neutrinos. Even though such sterile neutrinos have to direct standard model couplings they can still be produced in the early Universe through a combination of scattering and mixing. For certain masses and mixing angles sterile neutrinos can be completely thermalised in the early Universe leading to observable effect and potential conflict with existing data. At present there are tentative hints for the existence of sterile neutrinos of eV mass from a variety of short baseline neutrino experiments (see e.g.\ \cite{Kopp:2013vaa,Giunti:2012bc}). The preferred range of masses and mixing angles leads to almost complete thermalization and sterile neutrinos should be present with approximately the same number density as active states (see e.g.\ \cite{Hannestad:2012ky} for a recent discussion of sterile neutrino thermalisation). However, this is in disagreement with cosmological data analyses which find that eV mass sterile neutrinos may be allowed, but only provided their number density is significantly smaller than that of an active species. At first sight this seems to exclude the sterile neutrino hypothesis. However, there are significant loopholes in this argument because various physical mechanisms can prevent sterile neutrino production.
For example a large lepton asymmetry or new interactions in the sterile sector, as will be discussed below, can block the production of sterile neutrinos either completely or partially.

\subsection{Non-standard interactions}

A completely different type of beyond standard model physics in the neutrino sector arises in models with non-standard neutrino interactions. Non-standard interactions cover an extremely broad variety of models, but here I will simply discuss two
possibilities which are fairly representative of most models with additional interactions confined to the neutrino sector. The first is 
a four-point Fermi-like interaction induced by a new massive vector boson, $X$, and the second is a model in which neutrinos couple to a new light pseudoscalar particle such as the majoron.

In the 4-point interaction model neutrinos are strongly self-coupled at high temperature and subsequently decouple because $\Gamma/H \propto T^3$. However, the interaction can be significantly stronger than the standard model weak interaction because $X$ couples only to neutrinos. In \cite{Cyr-Racine:2013jua} cosmological bounds on this type of model were discussed and it was found that the 4-point interaction can actually be compatible with very high values of the effective coupling strength $G_X$ (corresponding to very a low mass of the mediator, $X$). However, there are significant bounds on this type of interactions from other sources such as neutrino scattering in the Sun \cite{Laha:2013xua}, and it is not completely clear whether the cosmologically allowed values of $G_X$ are already excluded by other data.

Another scenario is one where neutrinos interact with a new massless or very light pseudoscalar, as for example in majoron models. In this type of model neutrinos interact weakly early in the evolution of the Universe and subsequently become strongly self-interacting because $\Gamma/H \propto T^{-1}$. This means that such neutrino interactions can have a profound impact on structure formation because they can suppress neutrino anisotropic stress which in turn has a significant impact on the CMB spectrum. In fact the strongest known bound on light neutrino decays via pseudoscalar emission was derived using exactly this argument \cite{Hannestad:2005ex} (see also \cite{Bell:2005dr,Beacom:2004yd,Hannestad:2004qu}) for a discussion of this class of models.

\subsection{Sterile neutrinos {\it and} non-standard interactions?}

\begin{figure}[t]
\center
\includegraphics[height=.43\textwidth]{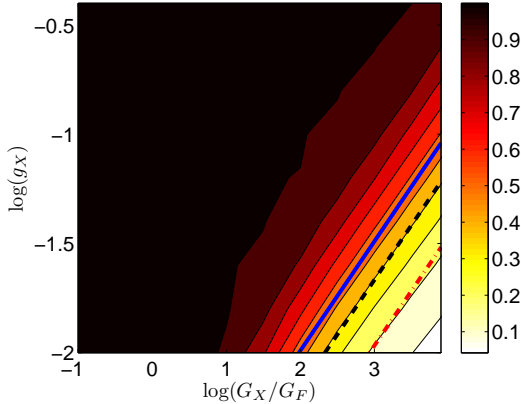}
\caption{
Contours of equal thermalization of the sterile neutrino. $\Delta N_\nu$ is given by the colors. The solid, dashed, and dot-dashed lines correspond to hidden bosons with masses $M_X = 300$, 200, and 100 MeV respectively (reproduced from \cite{Hannestad:2013ana}).
}
\label{fig:fig1}
\end{figure}

One of the problems with light sterile neutrinos in cosmology is that the masses and mixing angles indicated by short baseline experiments typically cause sterile neutrinos to be completely thermalised in the early universe. Since the mass is typically around 1 eV they will have too large an impact on structure formation to comfortably fit current observations and are therefore disfavoured to some extent. A possible and quite elegant loophole in this argument appears if sterile neutrinos are charged under a new gauge group with a light vector boson. In that case sterile neutrinos can generate a very strong matter potential for themselves and effective shut off further production either completely or partially. This leaves ample room for sterile neutrinos of eV mass to be compatible with current cosmological observations \cite{Hannestad:2013ana,Dasgupta:2013zpn}. In Fig.~\ref{fig:fig1} I show an example of how the degree of thermalisation can vary with model parameters for the new interaction (figure reproduced from \cite{Hannestad:2013ana}). Furthermore the light mediator needed in this model could well couple to dark matter and make it strongly self-interacting.

\section{Towards the future}

\begin{figure}[t]
\center
\includegraphics[height=.63\textwidth,angle=270]{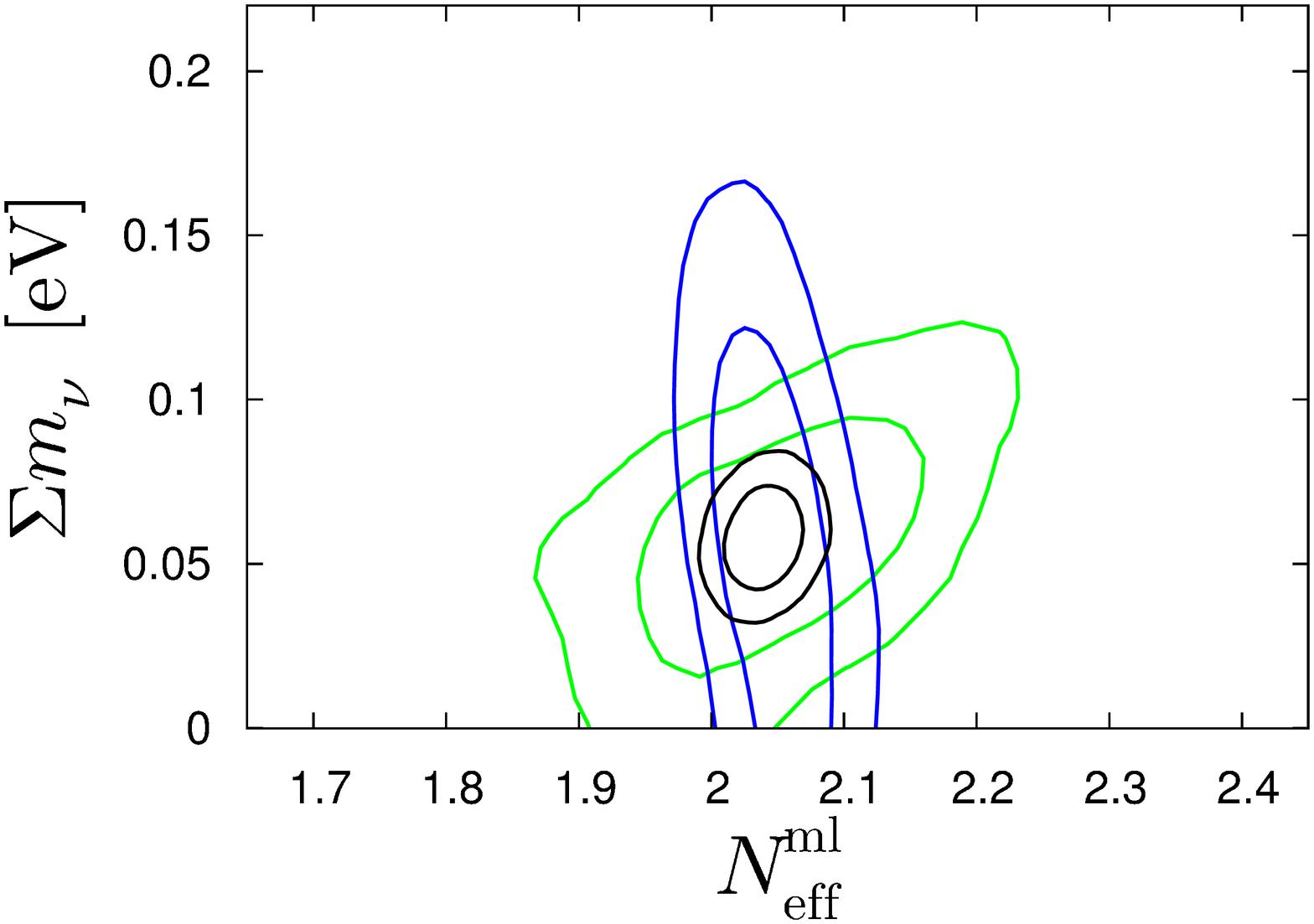}
\caption{
Marginalised joint two-dimensional  68\% and 95\% credible contours for $\sum m_\nu$ and the effective number of massless neutrinos from the EUCLID-like survey forecast performed in \cite{Basse:2013zua}. The different contours are for Planck CMB data + a EUCLID-like cluster data set (blue), Planck CMB data + a EUCLID-like photometric shear and galaxy survey (green),
and all data sets combined (black) (Reproduced from \cite{Basse:2013zua}).
}
\label{fig:fig2}
\end{figure}

The coming years will see big improvements in large scale surveys, with some surveys covering a substantial fraction of the current Hubble volume. For example the EUCLID sattelite will provide a photometric redshift and weak lensing survey of approximately 15,000 square degrees out to a redshift of approximately 2 \cite{Laureijs:2011gra}. Because large scale structure formation is exceedingly sensitive to effects of massive neutrinos such surveys will reach a sensitivity where a detection of the neutrino mass should be possible even for strongly hierarchical masses in the normal hierarchy ordering, i.e.\ the worst of all cases. The projected sensitivity is of order 0.01eV at 1$\sigma$, corresponding to approximately a 5$\sigma$ detection of the neutrino mass in this case \cite{Hamann:2012fe,Basse:2013zua}. An example of such a parameter forecast is shown in Fig.~\ref{fig:fig2} (figure reproduced from  \cite{Basse:2013zua}).

However, even with this extreme improvement a cosmological measurement of the mass ordering seems unlikely. To first order the suppression of power depends only on the total energy density in neutrinos, not on the distribution between mass eigenstates. Only in the region in $k$-space corresponding roughly to the horizon size when neutrinos go non-relativistic is there a pronounced difference between different hierarchies. Even with an effective volume as large as the one covered by EUCLID the sample variance is simply too large to see this effect. 

\section{Conclusions}

I have reviwed the current status of cosmological bounds on neutrino properties both within the standard model and in various extensions of it. The cosmological bound on the neutrino mass is currently in the range of 0.3-1 eV, significantly stronger than bounds from direct experiments, but much weaker than what is needed for a detection of neutrino masses is they are hierarchical. Future large scale surveys such as the one performed by the EUCLID satellite most likely will allow for the detection of a non-zero neutrino mass, even with hiearchical neutrino masses in the normal hierarchy. 
I have also discussed the current status of light sterile neutrinos in structure formation and how the cosmological bounds on sterile neutrinos fit together with current experimental results on sterile neutrinos from short baseline oscillation experiments. 
In summary, cosmology is an exccedingly powerful laboratory for doing neutrino physics and makes it possible to probe neutrino properties that are currently inaccessible to laboratory measurements.



%


\begin{thebibliography}{99}

\bibitem{Hannestad:2010kz} 
  S.~Hannestad,
  Prog.\ Part.\ Nucl.\ Phys.\  {\bf 65}, 185 (2010); 
  
	
	\bibitem{wong}
	Y.~Y.~Y.~Wong,
  Ann.\ Rev.\ Nucl.\ Part.\ Sci.\  {\bf 61}, 69 (2011);
  
	
	\bibitem{lesgourgues}
	J.~Lesgourgues and S.~Pastor,
  Adv.\ High Energy Phys.\  {\bf 2012}, 608515 (2012).
	
\bibitem{Ade:2013zuv}
  P.~A.~R.~Ade {\it et al.}  [Planck Collaboration],
  arXiv:1303.5076 [astro-ph.CO].
	
\bibitem{Hamann:2013iba}
  J.~Hamann and J.~Hasenkamp,
  JCAP {\bf 1310} (2013) 044
  [arXiv:1308.3255 [astro-ph.CO]].
	
\bibitem{Archidiacono:2013fha}
  M.~Archidiacono, E.~Giusarma, S.~Hannestad and O.~Mena,
  arXiv:1307.0637 [astro-ph.CO].
	
	%
\bibitem{Hannestad:2013ana}
  S.~Hannestad, R.~S.~Hansen and T.~Tram,
  arXiv:1310.5926 [astro-ph.CO].
	
\bibitem{Hannestad:2012ky}
  S.~Hannestad, I.~Tamborra and T.~Tram,
  JCAP {\bf 1207} (2012) 025
  [arXiv:1204.5861 [astro-ph.CO]].
	
\bibitem{Dasgupta:2013zpn}
  B.~Dasgupta and J.~Kopp,
  arXiv:1310.6337 [hep-ph].
	
\bibitem{Hamann:2012fe}
  J.~Hamann, S.~Hannestad and Y.~Y.~Y.~Wong,
  JCAP {\bf 1211} (2012) 052
	
\bibitem{Basse:2013zua} 
  T.~Basse, O.~E.~Bjaelde, J.~Hamann, S.~Hannestad and Y.~Y.~Y.~Wong,
  arXiv:1304.2321 [astro-ph.CO].
	
	
	%
\bibitem{Kopp:2013vaa} 
  J.~Kopp {\it et al.},
  JHEP {\bf 1305}, 050 (2013)
	
%
\bibitem{Giunti:2012bc} 
  C.~Giunti {\it et al.},
  Phys.\ Rev.\ D {\bf 87}, 013004 (2013)

\bibitem{Hannestad:2004qu}
  S.~Hannestad,
  JCAP {\bf 0502} (2005) 011
  [astro-ph/0411475].

\bibitem{Cyr-Racine:2013jua}
  F.~-Y.~Cyr-Racine and K.~Sigurdson,
  arXiv:1306.1536 [astro-ph.CO].
	
\bibitem{Hannestad:2005ex}
  S.~Hannestad and G.~Raffelt,
  Phys.\ Rev.\ D {\bf 72} (2005) 103514
  [hep-ph/0509278].
	
%
\bibitem{Bell:2005dr}
  N.~F.~Bell, E.~Pierpaoli and K.~Sigurdson,
  Phys.\ Rev.\ D {\bf 73} (2006) 063523
  [astro-ph/0511410].
	
\bibitem{Beacom:2004yd}
  J.~F.~Beacom, N.~F.~Bell and S.~Dodelson,
  Phys.\ Rev.\ Lett.\  {\bf 93} (2004) 121302
  [astro-ph/0404585].
	
\bibitem{Laha:2013xua}
  R.~Laha, B.~Dasgupta and J.~F.~Beacom,
  arXiv:1304.3460 [hep-ph].
	
\bibitem{Laureijs:2011gra}
  R.~Laureijs {\it et al.}  [EUCLID Collaboration],
  arXiv:1110.3193 [astro-ph.CO].
	
\end{thebibliography}
\end{document}